# Horizontal velocity field near the hot plate in turbulent natural convection

Vipin Koothur*, Baburaj A. Puthenveettil

*E-mail: vipink159@gmail.com

**Abstract**

*We study the velocity field in a horizontal (x-y) plane 1.5 mm above the hot plate in turbulent natural convection using PIV at a Rayleigh number $Ra_w=10^6$ and Prandtl number $Pr=5.2$. The plane of measurement is inside the velocity boundary layer estimated from the natural convection boundary layer equations[7] as well as inside the velocity boundary layer due to the large scale flow[2, 5]. The boundary layer comprises of line plumes with sinking fluid between them. The instantaneous velocity variation from the center of the sinking fluid to the line plumes is found to deviate with the classical Prandtl-Blasius laminar boundary layer profile, which is assumed to be the nature of boundary layer by the GL theory [2, 5]. Our results agree well with the natural convection boundary layer profile. The time averaged mean velocity variation deviates from both natural convection and Blasius type profiles as expected as it depends on the orientation of the line plumes. Our measurement result is a proof to the theory of the presence of a natural convection boundary layer on both sides of a line plume [10].*

## 1. Introduction

Turbulent natural convection is characterized by the following non dimensional numbers, Rayleigh Number $Ra_w=g\beta\Delta T_w H3/ (\alpha v)$; Prandtl Number $Pr=v/\alpha$; and aspect ratio $AR=D/H$. Here, $g$=acceleration due to gravity, $\beta$=the coefficient of thermal expansion, $\Delta T_w$=the temperature difference between the hot plate and the bulk, $H$=the fluid layer height, $v$=the kinematic viscosity, $\alpha$=the thermal diffusivity and $D$=the horizontal dimension of the fluid layer.

The regions close to the hot plate in turbulent natural convection consist of line plumes which merge with time [6, 7], surrounded by the entrainment flow between them. These together form a complex plume structure, which plays an important role in transfer of heat to the bulk and also in driving the large scale flow. Since the heat flux is decided by the boundary layers in between the plumes, it is important to understand the nature of such boundary layers.

All of the present and earlier theories on turbulent convection were based on certain boundary layer assumptions. The most prominent among them are the Grossmann and Lohse scaling theory [2, 5], where the boundary layer spanning the width of the cells is assumed to be laminar Prandtl-Blasius so as to estimate the boundary layer contributions to kinematic and thermal energy dissipation rates. Theerthan and Arakeri [10] and Puthenveettil et.al [6, 7], had assumed natural convection boundary layers between plumes, and based on the similarity solutions for natural convection boundary layers by Rotem and Claassen[8], had obtained near wall profiles, spacing and total length .Du puits et.al [3] from their work found deviations from Blasius profile due to the presence of thermal plumes detaching from the boundary layer. Zhou et.al [12] later using a method of dynamical rescaling concluded that the boundary layer had Blasius profile. Shi, Emran and Schumacher [9] from their DNS analysis of boundary layers showed that near-wall dynamics had a mixed convection profile.

In this work, we try to understand the structure of the local boundary layers on both sides of the plumes in turbulent natural convection by studying the horizontal profiles of the horizontal velocity inside the boundary layer and comparing it with classical Blasius and natural convection profile. Our study includes the analysis of both instantaneous as well as time averaged velocity profiles.

## 2. Theory

### *2.1. Prandtl-Blasius boundary layer*

In GL theory, Prandtl-Blasius boundary layer is assumed to be created by the large scale flow. The thickness of the boundary layer is given by the equation

$$\delta_{vb} = 0.482L/\sqrt{Re_L} \qquad (1)$$

where

$$Re_L = 0.204 Ra_w^{0.447} Pr^{-0.7} \qquad (2)$$

is the Reynolds number based on the fluid layer width and large scale velocity $U$ [2] given by

$$U = vRe_L/L \qquad (3)$$

The Prandtl-Blasius solution is given as

$$u_{bl} = U f'(\eta_{bl}) \quad (4)$$

where the similarity variable

$$\eta_{bl} = z\sqrt{Re_L}/x \quad (5)$$

A polynomial fit for the $f'(\eta_{bl})$ as a function of $\eta_{bl}$ from the data obtained from the numerical solutions of Blasius boundary layer equation (4) gives

$$f'(\eta_{bl}) = 0.00197\eta_{bl}^4 - 0.02397\eta_{bl}^3 + 0.049\eta_{bl}^2 + 0.306\eta_{bl} + 0.00132 \quad (6)$$

Substituting equation (6) in equation (4), gives the horizontal variation of horizontal velocity inside a Prandtl-Blasius boundary layer at a fixed height $z$, using equation (5).

## 2.2. Natural convection boundary layer

Each plume has a velocity boundary layer associated with it, with the thermal boundary layer nested inside it as shown in Figure 1. The thickness of this velocity boundary layer is given by the relation

$$\delta_v \sim (vL/u_c)^{0.5} \quad (7)$$

where $L$ is the characteristic length scale (half the mean plume spacing $\lambda$) and $u_c$ is the characteristic velocity obtained by a balance of buoyancy driven pressure gradient and viscous resistance [7].

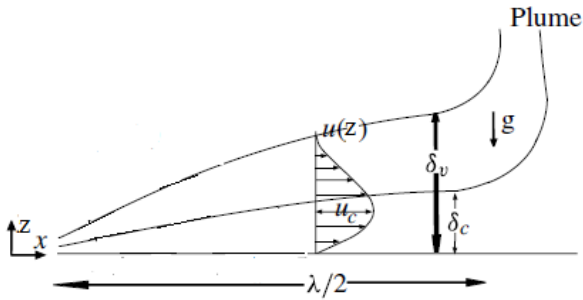

Figure 1. Schematic of laminar natural convection boundary layer giving rise to a plume.

From the similarity solutions of Rotem and Claassen [8] for natural convection boundary layers over a horizontal isothermal plate, the relation for horizontal velocity inside the boundary layer is given as

$$u_{nc} = vGr_L^{2/5} f'(\eta_{nc}) x^{1/5}/L, \quad (8)$$

Where

$$Gr_L = g\beta\Delta T_w L^3/v^2 \quad (9)$$

is the Grashoff number based on $\Delta T_w$ and $L$ and similarity variable

$$\eta_{nc} = zGr_L^{1/5} x^{-2/5} \quad (10)$$

A polynomial fit of $f'(\eta_{nc})$ from the data provided by Rotem and Claassen [8] gives $f'(\eta_{nc})$ as a function of $\eta_{nc}$ at $Pr=5.2$ as

$$f'(\eta_{nc}) = 0.0005516\eta^5 - 0.01122\eta^4 + 0.08996\eta^3 - 0.344\eta^2 + 0.5391\eta + 0.007895 \quad (11)$$

Substituting this in equation (8) we get the horizontal velocity as a function of horizontal distance $x$, at any fixed height $z$ inside a natural convection boundary layer, using equation (10).

## 3. Experiment

### 3.1. Setup and Procedure

An open tank of size 300×300×250 mm, with four glass side walls was used in the experiment. The bottom hot plate was 10 mm thick and was made of copper. The heat was supplied from resistance heating of a Ni-Cr wire placed below an aluminum plate. The Al plate was separated from the Cu plate by a glass plate. The temperature difference measured across the glass plate gives the heat flux supplied to the copper plate.

The experiment was conducted at $Ra=3.96\times10^6$, $Pr=5.2$ and $AR=6$. The technique of particle image velocimetry (PIV) was used to measure the velocity fields in a horizontal plane 1.5 mm above the hot plate. The measurement plane was well inside both boundary layer thicknesses obtained from equation (1) and (6). The selected area of imaging was 80×80 mm$^2$ corresponding to 64×64 velocity vectors, at the center of the hot plate.

### 3.2. 2D PIV measurement

The flow was seeded with polyamide spheres of size 55 μm and specific gravity of 1.016 g/cm$^3$. These particles were illuminated from a Nd:YAG laser (Litron) with a laser sheet of thickness of 1 mm. The in-plane and out of plane velocities were estimated to find the time difference between the laser pulses, using the relations for horizontal and vertical velocities for a plume as given in [4]. Single pulse single frame technique was employed to capture the images at a frequency of 5Hz, using a Pro X CCD camera (2048×2048 pixels). The camera was aligned normal to the plane of measurement. The velocity fields were obtained using cross-correlation processing in time series mode, with an interrogation window size of 32×32 pixels with 50% overlap, using Davis 7.2 software.

## 4. Result and Discussion

### 4.1. Instantaneous velocity field analysis

Line plumes were identified from the velocity vector plot by taking the divergence of the velocity field [11]. Figure 2a. shows an instantaneous velocity field, overlaid over the divergence of the velocity field. The regions with negative divergence show the rising plumes and the regions with positive divergence show the sinking fluid in between the plumes.

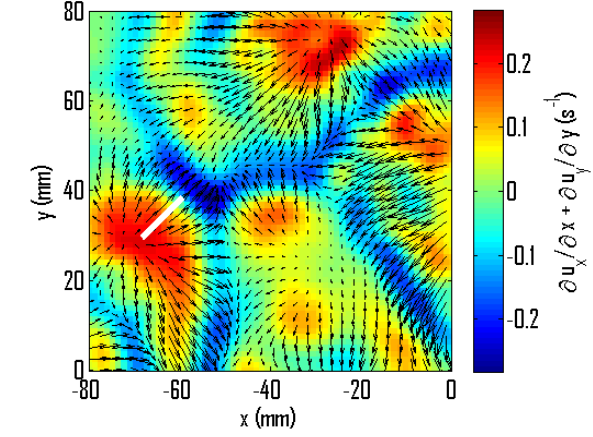
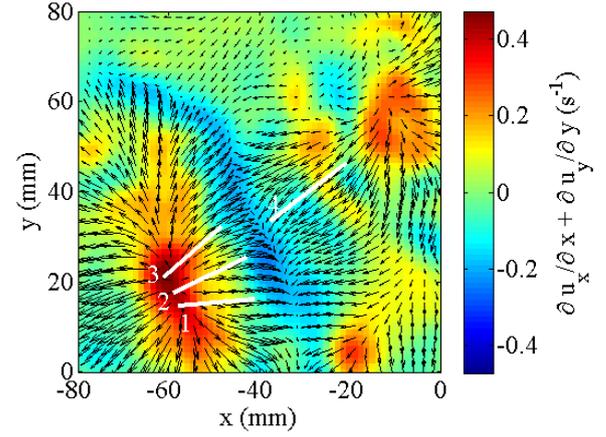
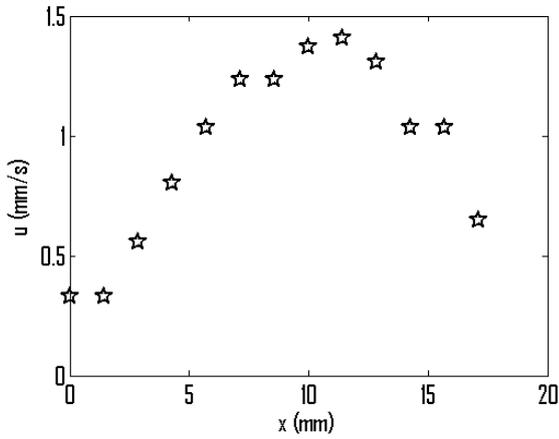
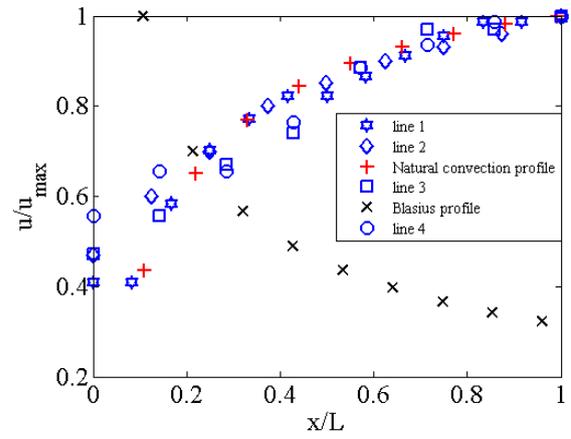

Figure 2 (a) Instantaneous velocity field overlaid over the divergence of the velocity, a straight line is taken perpendicular to the plume at a location (b) Variation of velocity $u=\sqrt{u_x^2+u_y^2}$ with distance x along the line taken in (a)

Figure 3 (a) Instantaneous velocity field overlaid over the divergence field after the plumes in Fig 2a. have merged (b)Comparison of experimental result with the natural convection and Blasius profile.

A straight line perpendicular to the plumes is drawn from the center of the region with sinking fluid to the edge of the line plumes at a location where the velocity vectors are along the line, as shown in Figure 2a. The variation of instantaneous horizontal velocity $u=\sqrt{(u_x^2+u_y^2)}$ along this straight line is shown in Figure 2b. The horizontal velocity increases from the center of the sinking fluid and reaches maximum $u_{max}$ at a certain distance and then decreases until the edge of the plume. Since the decreasing part of the velocity field is due to the stagnation just before the flow turns upward corresponds to flow into the plumes, we neglect this and define the instantaneous characteristic length scale (L) as the distance from the center of the sinking fluid to the point of maximum instantaneous velocity.

Figure 3a. shows a vector field after Figure 2a where the line plumes in Figure 2a have merged. Horizontal velocity u is normalized by the maximum instantaneous velocity $u_{max}$ and the horizontal distance x is normalized by the characteristic length scale L at different locations from Figure 3a. Figure 3b. shows comparison of the horizontal profile of the horizontal velocity obtained from the experiment with that obtained from equation (4) for classical Prandtl-Blasius boundary layer and equation (7) for natural convection boundary layer. The experimental results agree well with the natural convection boundary layer profile, while the Prandtl-Blasius profile has a completely different trend from the experimental trend.

### 4.1. Analysis of spatio-temporal averaged velocity field

The spatio-temporal averaged velocity is given as

$$u(x,y)=\langle\langle u(x,y,t)\rangle_{x,y}\rangle_t \qquad (10)$$

where $\langle\ldots\rangle_t$ denoting the ensemble average was calculated by time averaging over a period of one large scale circulation and $\langle u \rangle_x$ and $\langle u \rangle_y$ by spatial averaging along x and y size of the image. The time for one large scale circulation for this experiment was 11s with the large scale velocity $U$=4.7mm/s obtained from equation (3). Figure 4a. shows time averaged velocity fields, spatial mean of the ensemble averaged velocity field were taken along x and y directions to get ⟨⟨u⟩$_x$⟩$_t$ and ⟨⟨u⟩$_y$⟩$_t$.

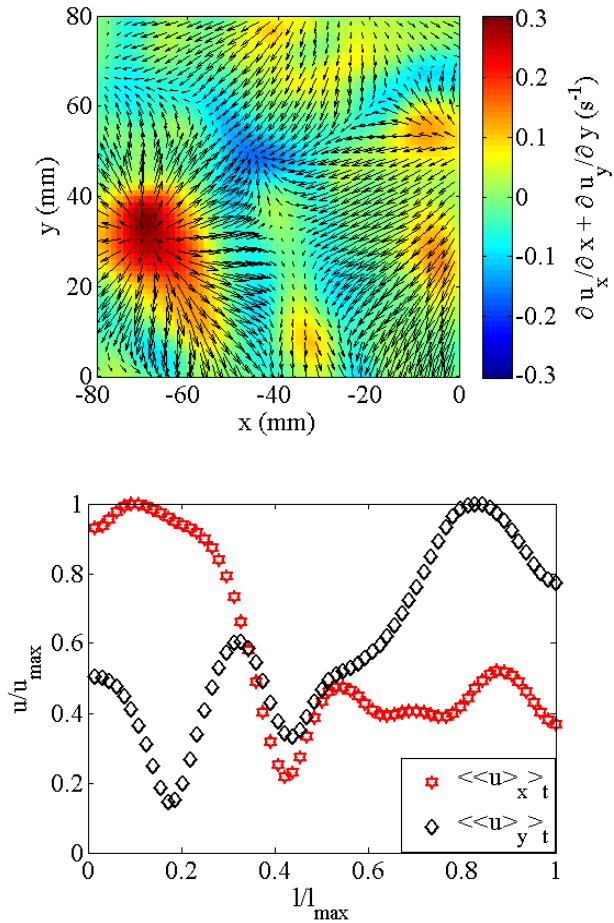

Figure 4 (a) Temporal averaged vector field overlaid over divergence of the velocity field (b) Variation of spatio-temporal averaged velocity along x and y with the respective distance.

Figure 4b. shows variation of ⟨⟨u⟩$_x$⟩$_t$ and ⟨⟨u⟩$_y$⟩$_t$ for Figure 4a. As expected they deviate from both the natural convection boundary layer and Blasius boundary layer profiles. This is because they are dependent on the plume orientation. The profiles will be different for different plume orientations.

## Conclusion

In this paper, we compared the horizontal profile of the horizontal velocity in a natural convection boundary layer and classical Prandtl-Blasius boundary layer with the experimental profile of horizontal velocities in between the plumes. The time averaged profile does not give any physical meaning and are dependent on the plume orientations whereas the instantaneous profiles are in good agreement with natural convection boundary layer profile but deviates from the classical Prandtl-Blasius boundary layer profile.

*Vipin Koothur, Department of Applied Mechanics, Indian Institute of Technology Madras, Chennai, India.*

*Dr. Baburaj A. Puthenveettil, Department of Applied Mechanics, Indian Institute of Technology Madras, Chennai, India.*